\documentclass{PoS}

\newcommand\PSb{\Phi_B}

\newcommand\POWHEG{{\tt POWHEG}}

\newcommand\MiNLO{{\tt MiNLO}}
\def\sss{\scriptscriptstyle}
\newcommand\as{\alpha_{\sss\rm S}}

\newcommand\Phirad{\Phi_{\mathrm{rad}}}

\newcommand{\Matrix}{\scalebox{1.0}{\textsc{Matrix}}}
\newcommand{\Munich}{\scalebox{1.0}{\textsc{Munich}}}

\newcommand{\OpenLoops}{{\sc OpenLoops}}

\newcommand{\yww}{\ensuremath{y_{WW}}}
\newcommand{\dywpwm}{\ensuremath{\Delta y_{W^+W^-}}}

\newcommand{\ptwm}{\ensuremath{p_{T,W^-}}}

\newcommand{\mwp}{\ensuremath{m_{W^+}}}
\newcommand{\mwm}{\ensuremath{m_{W^-}}}
\newcommand{\thetap}{\ensuremath{\theta^{\rm\scalebox{0.6}{CS}}_{W^+}}}
\newcommand{\costhetap}{\ensuremath{{\rm cos}\,\thetap}}
\newcommand{\phip}{\ensuremath{\phi^{\rm\scalebox{0.6}{CS}}_{W^+}}}
\newcommand{\thetam}{\ensuremath{\theta^{\rm\scalebox{0.6}{CS}}_{W^-}}}
\newcommand{\costhetam}{\ensuremath{{\rm cos}\,\thetam}}
\newcommand{\phim}{\ensuremath{\phi^{\rm\scalebox{0.6}{CS}}_{W^-}}}

\newcommand{\thetacs}{\theta^*}
\newcommand{\phics}{\phi^*}

\newcommand{\dd}{{\rm d}}
\newcommand{\dptwm}{\ensuremath{\dd\ptwm}}
\newcommand{\dyww}{\ensuremath{\dd\yww}}
\newcommand{\ddywpwm}{\ensuremath{\dd\dywpwm}}

\newcommand{\dmwp}{\ensuremath{\dd\mwp}}
\newcommand{\dmwm}{\ensuremath{\dd\mwm}}
\newcommand{\dcosthetap}{\ensuremath{\dd\costhetap}}
\newcommand{\dphip}{\ensuremath{\dd\phip}}
\newcommand{\dcosthetam}{\ensuremath{\dd\costhetam}}
\newcommand{\dphim}{\ensuremath{\dd\phim}}
\newcommand{\ptjetveto}{\ensuremath{p_{T,j_1}^{\rm veto}}}
\newcommand{\POWHEGBOX}{{POWHEG BOX}}
\newcommand{\ptjetone}{\ensuremath{p_{T,j_1}}}

\newcommand{\ptll}{\ensuremath{p_{T,\ell\ell}}}
\newcommand{\ptmiss}{\ensuremath{p_{T}^{\textrm{\scriptsize miss}}}}
\def\nn{\nonumber}

\newcommand{\D}{\mathrm{d}}

\title{$W^+W^-$ production at NNLO+PS}

\ShortTitle{$W^+W^-$ production at NNLO+PS}

\author{
  \speaker{Emanuele Re}
  \thanks{This research project has been supported by a
    Maria Sk\l{}odowska-Curie Individual Fellowship of the European
    Commission's Horizon 2020 Programme under contract number 659147
    PrecisionTools4LHC. This work was also supported in part by the
    ERC Consolidator Grant HICCUP (No. 614577).
    This work has been performed with the collaboration of
    Marius Wiesemann and Giulia Zanderighi.}~\thanks{Preprint
    numbers: CERN-TH-2018-192, LAPTH-Conf-032/18}\\~\\ CERN,
  Theoretical Physics Department, 1211 Geneva 23, Switzerland\\ LAPTh,
  CNRS, Universit\'e Savoie Mont Blanc, 74940 Annecy, France\\ E-mail:
  \email{emanuele.re@lapth.cnrs.fr}}

\abstract{We present predictions for $W^+W^-$ production (with exact
  decays) that are next-to-next-to-leading order (NNLO) accurate and
  consistently matched to a parton shower. The matching is achieved by
  upgrading, with \MiNLO{}, the NLO calculation of $W^+W^-$+1 jet
  production, in such a way that NLO accuracy is guaranteed for
  $W^+W^-$ inclusive observables, and by performing subsequently a
  reweighting of the {\tt WWJ-MiNLO} events, differential in the
  $W^+W^-$ Born variables, to the NNLO results obtained with
  \Matrix{}.}

\FullConference{XXVI International Workshop on Deep-Inelastic Scattering and Related Subjects (DIS2018)\\
		16-20 April 2018\\
		Kobe, Japan}

\begin{document}

The study of vector boson pair-production is central for the LHC
Physics program. Not only is $W^+W^-$ production measured to probe
anomalous gauge couplings, but it's also an important background for
several searches, notably for those where the $H\to W^+W^-$ decay is
present. For these and other similar reasons, it is important to have
flexible and fully differential theoretical predictions that allow to
model simultaneously, and at least with QCD NLO accuracy, both
inclusive $W^+W^-$ production, as well as $W^+W^-$ production in
presence of jets. Methods aiming at this task are usually referred to
as ``NLO+PS merging''. NLO+PS merging for $pp\to VV$+jet(s) was
achieved using the {\ttfamily
  MEPS@NLO}~\cite{Hoeche:2012yf,Cascioli:2013gfa}, {\ttfamily
  FxFx}~\cite{Frederix:2012ps,Alwall:2014hca}, and
\MiNLO{}~\cite{Hamilton:2012rf,Hamilton:2016bfu} methods.

Although NLO accuracy is indispensable, in the context of searches
involving a pair of gauge bosons the experimental precision reached by
the LHC experiments already demands for predictions whose accuracy
goes beyond NLO(+PS). In fact, for inclusive and fiducial total cross
section, comparing data with NNLO-accurate results is already crucial,
and, sooner rather than later, this will be the case also for
differential measurements. Therefore matching NNLO results to parton
showers (NNLO+PS) for diboson production is an important goal, as it
allows to combine, in a single and flexible simulation, higher-order
(fixed-order) corrections with the benefits of an all-order
description (as given by the parton shower), which is important
in some corners of the phase space.

In this manuscript we'll describe the results
presented in ref.~\cite{Re:2018vac}, where the NNLO QCD corrections
for $W^+W^-$ production in hadron collision\footnote{Throughout this
  document, we use the shortcut notation ``$W^+W^-$ production'' to
  actually denote the full process $pp\to {e}^-\bar{\nu}_e
  \mu^+{\nu}_\mu$, \emph{i.e.} $W$ bosons are treated as unstable,
  have a finite width and their leptonic decay is treated exactly in
  all the matrix elements used. When relevant, we'll describe the
  approximation we made.} were consistently matched to parton showers,
thereby obtaining, for the first-time, a NNLO+PS accurate simulation
of diboson production. In section~\ref{sec:minlo_nnlo}, we'll describe
the two inputs that we used, namely the \MiNLO{} method used to merge
at NLO+PS the $W^+W^-$ and $W^+W^-$+1 jet production processes, and
then, subsequently, the fully-differential NNLO computation of
$W^+W^-$ production using the \Matrix{} framework. In
section~\ref{sec:nnlops} instead the method and approximations we used
to upgrade the \MiNLO{} results to full NNLO accuracy will be
described, and show some validation and phenomenological results,
taken from ref.~\cite{Re:2018vac}.

\section{$W$-boson pair production: \MiNLO{} merging and NNLO results}
\label{sec:minlo_nnlo}

\subsection{{$W^+W^-+1$ jet NLO+PS merging with \MiNLO{}}}
The \MiNLO{} (Multi-scale Improved NLO)
procedure~\cite{Hamilton:2012np} was originally introduced as a
prescription to a-priori choose the renormalization ($\mu_R$) and
factorization ($\mu_F$) scales in multileg NLO computations: since
these computations can probe kinematic regimes involving several
different scales, the choice of $\mu_R$ and $\mu_F$ is indeed
ambiguous, and the \MiNLO{} method addresses this issue by
consistently including CKKW-like
corrections~\cite{Catani:2001cc,Lonnblad:2001iq} into a standard NLO
computation. In practice this is achieved by associating a
``most-probable'' branching history to each kinematic configuration,
through which it becomes possible to evaluate the couplings at the
branching scales, as well as to include (\MiNLO{}) Sudakov form
factors (FF). This prescription regularizes the NLO computation also
in the regions where jets become unresolved, hence the \MiNLO{}
procedure can be used within the \POWHEG{} formalism to regulate the
$\bar{B}$ function for processes involving jets at LO.

In a single equation, for a $q\bar{q}$-induced process as $W^+W^-$
production, the \MiNLO{}-improved \POWHEG{} $\bar{B}$ function reads:
\begin{equation}
  \label{eq:ww-minlo}
  \bar{B}_{\,\tt WWJ-MiNLO} =  \as(q_T) \Delta^2_q(q_T,M_X)   %\\
  \Big{[}
  B ( 1-2\Delta^{(1)}_q(q_T,M_X) ) + 
  \as V(\bar{\mu}_R) +\as \int \dd\Phirad R
  \Big{]}\nn\,,
\end{equation}
where $X$ is the color-singlet system ($W^+W^-$ in this case), $q_T$ is
its transverse momentum, $\bar{\mu}_R$ is set to $q_T$, and
$\Delta_q(q_T,Q)=\exp\Big{\{}-\int_{q^2_T}^{Q^2}\frac{\dd q^2}{q^2}\frac{\as(q^2)}{2\pi}
\Big{[} A_{q}\log\frac{Q^2}{q^2} + B_{q} \Big{]}\Big{\}}$ is the
\MiNLO{} Sudakov FF associated to the jet present at LO. Convolutions
with PDFs are understood, $B$ is the leading-order matrix element for
the process $pp\to X+1$ jet (stripped off of the strong coupling), and
$\Delta_{q}^{(1)}(q_T,Q)$ (the $\mathcal{O} (\as)$ expansion of
$\Delta_q$) is removed to avoid double counting.
%We also notice that
%$\bar{B}_{\,\tt WWJ-MiNLO}$ is a function of $\Phi_{X+j}$,
%\emph{i.e.} the phase space to produce the $X$ system and an extra
%parton, which can be arbitrarily soft and/or collinear.

In ref.~\cite{Hamilton:2012rf} it was also realized that, if $X$ is a
color singlet, upon integration over the full phase space for the
leading jet, one can formally recover NLO+PS accuracy for the process
$pp\to X$ by properly applying \MiNLO{} to NLO+PS simulations for
processes of the type $pp\to X+1$ jet.\footnote{The idea has been
  significantly extended in ref.~\cite{Frederix:2015fyz} to also treat
  processes beyond color-singlet production, and also applied,
  recently, to single-top production~\cite{Carrazza:2018mix}.}  In the
following we denote {\tt XJ-MiNLO'} a simulation with this property.
Besides setting $\mu_F$ and $\mu_R$ equal to $q_T$ in all their
occurrences, the key point is to include at least part of the
Next-to-Next-to-Leading Logarithmic (NNLL) corrections into the
\MiNLO{} Sudakov form factor, namely the $B_2$ term: by omitting it,
the full integral of eq.~(\ref{eq:ww-minlo}) over $\Phi_{X+j}$, albeit
finite, differs from $\sigma_{pp\to X}^{NLO}$ by a relative amount
$\as(M_X)^{3/2}$, thereby hampering a claim of NLO accuracy.

The $B_2$ coefficient is process-dependent, and formally also a
function of $\Phi_{X}$, because part of it stems from the one-loop
correction to the $pp\to X$ process. For Higgs, Drell-Yan, and $VH$
production, these one-loop corrections can be expressed as form
factors: $B_2$ becomes just a number as its dependence upon $\Phi_{X}$
disappears, and the analogous of eq.~(\ref{eq:ww-minlo}) can be easily
implemented~\cite{Hamilton:2012rf,Luisoni:2013kna}. For diboson
production, the extraction of $B_2$ is more subtle, as there's an
explicit dependence upon $\Phi_{X}$ which needs to be retained. To
deal with this issue, in ref.~\cite{Hamilton:2016bfu} we defined, on
an event-by-event basis, a projection of the $W^+W^-+1$ jet state onto
a $W^+W^-$ one, with the requirement that the $q_T\to 0$ limit is
approached smoothly. By combining the above points it was possible to
upgraded with \MiNLO{} a \POWHEG{} generator for the $pp\to W^+W^-+1$
jet process, thereby obtaining a NLO+PS merging for $pp\to W^+W^-$ and
$pp\to W^+W^-$+1 jet~\cite{Hamilton:2016bfu}.\footnote{Tree-level matrix elements were
  obtained with an interface to {\tt
    MadGraph\,4}~\cite{Alwall:2007st,Campbell:2012am}, whereas one
  loop corrections were computed with {\tt
    GoSam\,2.0}~\cite{Luisoni:2013kna,Cullen:2014yla}. We have worked
  in the 4-flavour scheme.} This generator, henceforth denoted as {\tt
  WWJ-MiNLO'}, is the one we used to produce the events that were then
upgraded to NNLO, as described in section~\ref{sec:nnlops}.

\subsection{{$W^+W^-$ production at NNLO}}
The computation of QCD NNLO corrections for differential cross
sections at the LHC has witnessed an enormous progress in the last few
years. Part of this progress is due to the variety and flexibility of
methods to handle the cancellation of collinear and infrared
divergences between the different NNLO terms of the computation. As a
result, essentially all processes with 2 hard objects (massless or
massive, and possibly decaying) in the final state that are relevant
for LHC phenomenology are now known at NNLO.

Among the available methods to compute NNLO corrections, the
``$q_T$-subtraction'' formalism~\cite{Catani:2007vq} has been widely
used for processes involving color-singlet production.
%Although
%there's ongoing work to extend it to also deal with more complicated
%processes~\cite{Bonciani:2015sha}, here we'll restrict the discussion
%to the production of a color-singlet final state $X$.
The main idea relies on the observation that, at NNLO, a generic
differential cross section for $pp\to X$ production can be written,
schematically, as
\begin{equation}
  \label{eq:qtsubtraction}
  \D{\sigma}^{X}_{\mathrm{NNLO}}={\cal H}^{X}_{\mathrm{NNLO}}\otimes \D{\sigma}^{X}_{\mathrm{LO}}
  +\left[ \D{\sigma}^{{X + jet}}_{\mathrm{NLO}}-
    \D{\sigma}^{\mathrm{CT}}_{\mathrm{NNLO}}\right]\,, \nn
\end{equation}
where $\D{\sigma}^{{X + jet}}_{\mathrm{NLO}}$ denotes the complete NLO
computation for $X+1$ jet, where all infrared and collinear
singularities associated to the $X+2$ partons matrix elements have
been properly regularized, \emph{i.e.} $\D{\sigma}^{{X +
    jet}}_{\mathrm{NLO}}$ is singular only when $q_T\to
0$. Eq.~(\ref{eq:qtsubtraction}) shows that, starting from
$\D{\sigma}^{{X + jet}}_{\mathrm{NLO}}$, one can compute the NNLO
corrections by regularizing the $q_T\to 0$ limit through the
counterterm $\D{\sigma}^{\mathrm{CT}}_{\mathrm{NNLO}}$ (which depends
on the process at hand only through the $pp\to X$ LO matrix elements,
and it can be built from resummed results for $p_T$ spectra), and
adding, separately, a ``hard-collinear'' function ${\cal
  H}^{X}_{\mathrm{NNLO}}$, which also contains the two-loop amplitudes.

As it will be explained in section~\ref{sec:nnlops}, in our work we
needed differential distributions for $W^+W^-$ production at NNLO. We
have obtained them using
\Matrix{}~\cite{Grazzini:2017mhc},\footnote{https://matrix.hepforge.org/}
which is a computational framework where a fully general
implementation of the $q_T$-subtraction formalism has been
implemented, and used to obtain NNLO QCD corrections to a large number
of hadron-collider processes with color-neutral final states. Of
particular interest for the case at hand are the results presented in
refs.~\cite{Gehrmann:2014fva,Grazzini:2016ctr}.\footnote{\Matrix{} has
  been used also in the NNLO diboson computations of
  refs.~\cite{Grazzini:2013bna,Grazzini:2015nwa,Cascioli:2014yka,Grazzini:2015hta,Grazzini:2016swo,Grazzini:2017ckn}
  and in computation of the NNLL+NNLO resummed $ZZ$ and $WW$ spectrum
  of ref.~\cite{Grazzini:2015wpa}.}

\Matrix{} makes use of an automated implementation of the
Catani-Seymour dipole subtraction
method~\cite{Catani:1996jh,Catani:1996vz} within the Monte Carlo
program \Munich{}.\footnote{\Munich{} is the abbreviation of
  ``MUlti-chaNnel Integrator at Swiss (CH) precision''---an automated
  parton level NLO generator by S.~Kallweit.} Moreover all (spin- and
colour-correlated) tree-level and one-loop amplitudes are obtained
from \OpenLoops{}~\cite{Cascioli:2011va,Buccioni:2017yxi}, whereas,
for the two-loop amplitudes, dedicated computations are employed: for
the process at hand, the two-loop amplitudes for the production of a
pair of off-shell massive vector bosons~\cite{Gehrmann:2015ora} are
taken from the publicly available code
%\textsc{VVamp}.\footnote{https://vvamp.hepforge.org/}

\section{$W$-boson pair production at NNLO+PS}
\label{sec:nnlops}
As first discussed in ref.~\cite{Hamilton:2012rf}, and then further
elaborated upon in ref.~\cite{Hamilton:2013fea}, it is possible to match a
parton shower simulation to a NNLO computation for $pp\to X$
production by means of a multi-differential reweighting, applied on an
event-by-event basis to a {\tt XJ-MiNLO'} simulation: the weight of
each {\tt XJ-MiNLO'} event has to be rescaled using the reweighting factor
${\cal W}(\PSb)$, defined as
\begin{equation}
  {\cal W}(\PSb) = \frac{\dd\sigma^{\mathrm{NNLO}}/\dd\PSb}{\dd\sigma^{\tt MiNLO}/\dd\PSb}\,,
\end{equation}
where $\PSb$ identifies the phase space for $pp\to X$
production.\footnote{In practice the reweighting procedure actually
  used is a bit more complicated, but in this manuscript we restrict
  our discussion to the simpler case, as it is enough to illustrate
  all the relevant points discussed.}  This procedure obviously gives
back results that are, by construction, NNLO accurate for inclusive
observables (\emph{i.e.} observables that only depend upon
$\PSb$). Moreover, since formally ${\cal W}(\PSb)=1+\mathcal{O}(\as^2)$, the
NLO accuracy of the $X+1$ jet phase space region is not spoiled.

The above procedure has been applied successfully for Higgs ($pp\to
H$)~\cite{Hamilton:2013fea}, Drell-Yan ($pp\to Z\to
\ell\bar{\ell}$)~\cite{Karlberg:2014qua} and $VH$ production ($pp\to
\ell\bar{\ell}H$)~\cite{Astill:2016hpa,Astill:2018ivh}. Nevertheless,
since multi-differential distribution at NNLO are required to compute
${\cal W}(\PSb)$, it is easy to realize that the higher is the
dimensionality of the $\PSb$ phase space, the more challenging the
method becomes: for Higgs, Drell-Yan and $VH$ production, the
dimensionality of the final state phase space is one, three and six,
respectively. For diboson production ($pp\to {e}^-\bar{\nu}_e
\mu^+{\nu}_\mu$) the Born phase space becomes 9-dimensional.

Computing a differential cross section at NNLO as a function of 9
observables is at the moment numerically unfeasible. In the following we
quickly describe how we handled the computational complexity in the
diboson case: after having chosen to parametrize $\PSb$ in terms of 9 variables,
\begin{equation}
  \dd\PSb={\dptwm\dyww\ddywpwm\dcosthetap\dphip\dcosthetam\dphim\dmwp\dmwm}\,,
\end{equation}
where $\theta^{CS}_{W^\pm}$ and $\phi^{CS}_{W^\pm}$ are the
Collins-Soper angles~\cite{Collins:1977iv} for the ${e}^-\bar{\nu}_e$ and $\mu^+{\nu}_\mu$
leptonic pairs, $\mwm$ and $\mwp$ are their respective invariant
masses, $\ptwm$ is the $W^-$-boson transverse momentum, $\yww$ the
rapidity of the diboson pair, and $\dywpwm$ the rapidity difference
between the two $W$ bosons, we have made the following approximations
to compute ${\cal W}(\PSb)$:
\begin{itemize}
\item[-] we have dropped the dependence of the differential cross
  section upon the $\mwm$ and $\mwp$ virtualities, as the expectation
  is that the NNLO-to-NLO K-factor is flat in these 2 directions;
\item[-] despite being strictly speaking correct only for the
  double-resonant topologies, for each $W$-boson decay, we
  parametrized the functional dependence of the cross-section upon the
  Collins-Soper angles using 9 functions $f_i(\thetacs,\phics)$,
  $i=0,...,8$. These functions are well-known combinations of the 9
  spherical harmonics $Y_{lm}(\thetacs,\phics)$, $l\le 2$ and $|m|\le
  l$~\cite{Collins:1977iv}.
\end{itemize}
After implementing the above approximations, the expression for the
fully differential cross section becomes
\begin{equation}
  \label{eq:simple}
  \frac{\dd\sigma}{\dd\PSb}=\frac{9}{256\pi^2}\,\sum\limits_{i=0}^8 \sum\limits_{j=0}^8 
AB_{ij}\,f_i(\thetam,\phim)\,f_j(\thetap,\phip)\,.
\end{equation}
The 81 coefficients $AB_{ij}$ are functions of the 3 remaining
variables $(\ptwm, \yww, \dywpwm)$, and, since the set $\{f_i
f_j\}_{i,j=0,...,8}$ is a basis, they can be obtained by suitable
projections. From eq.~(\ref{eq:simple}) it is clear that we have
traded the task of computing a 9-dimensional NNLO differential cross
section with that of extracting the 81 triple-differential
distributions $AB_{ij}(\ptwm, \yww, \dywpwm)$. Although the task is
still numerically challenging, in ref.~\cite{Re:2018vac} we have used
this procedure to compute ${\dd\sigma^{\mathrm{NNLO}}/\dd\PSb}$ (from
\Matrix{}) and $ {\dd\sigma^{\tt MiNLO}/\dd\PSb}$ (from {\tt
  WWJ-MiNLO'}) and, in turn, to reweight the \MiNLO{}
events to NNLO. In the rest of this section we report some of the
results shown in ref.~\cite{Re:2018vac}, where all the technical
choices and setting of parameters not described here can be
found. Here we only stress that, in our results, we have not included
the loop-induced contribution $gg\to W^+W^-$, which enters first at
NNLO, and whose size is about 30\% of the remaining
$\mathcal{O}(\as^2)$ corrections.

\begin{figure}
  \includegraphics[width=0.49\textwidth]{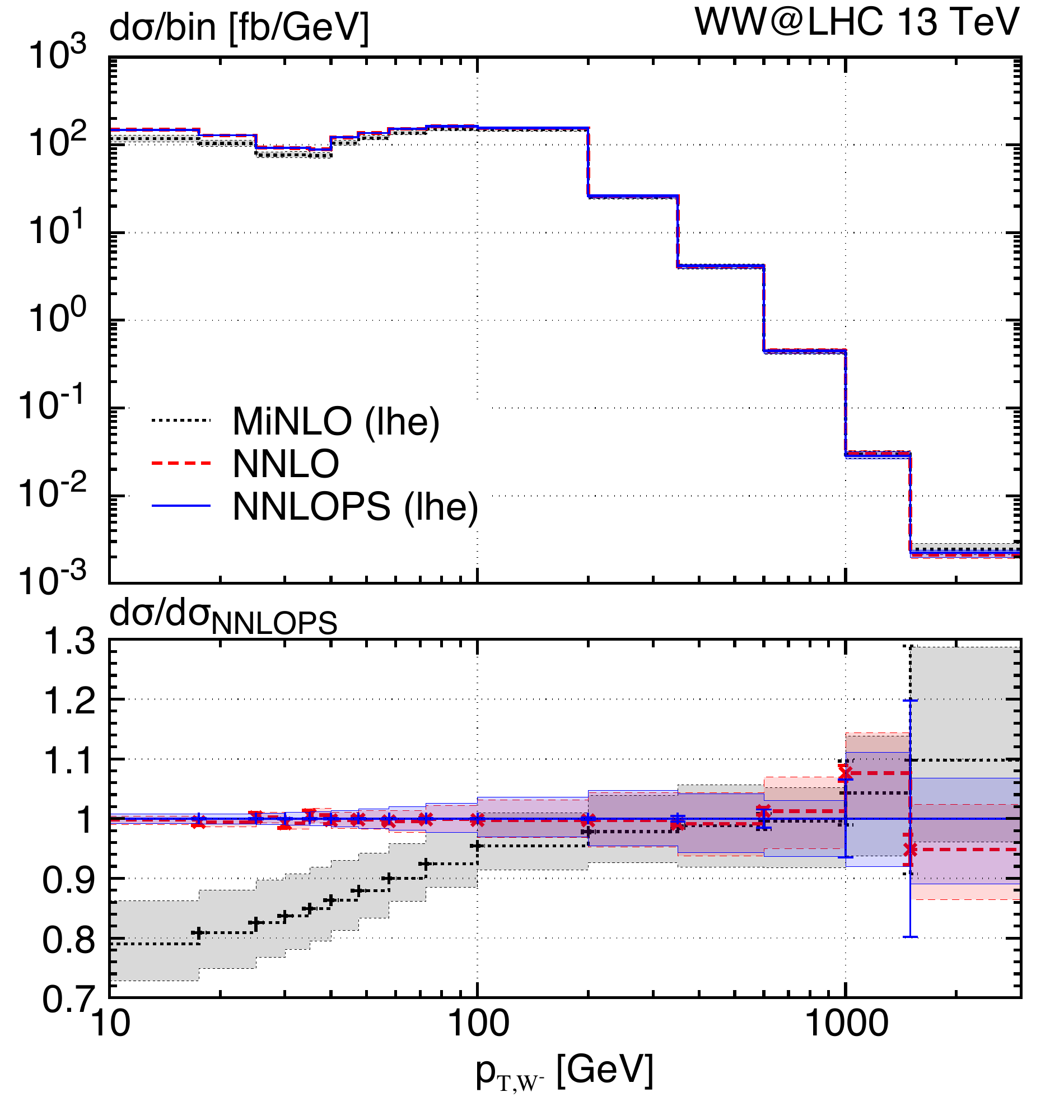}~
  \includegraphics[width=0.49\textwidth]{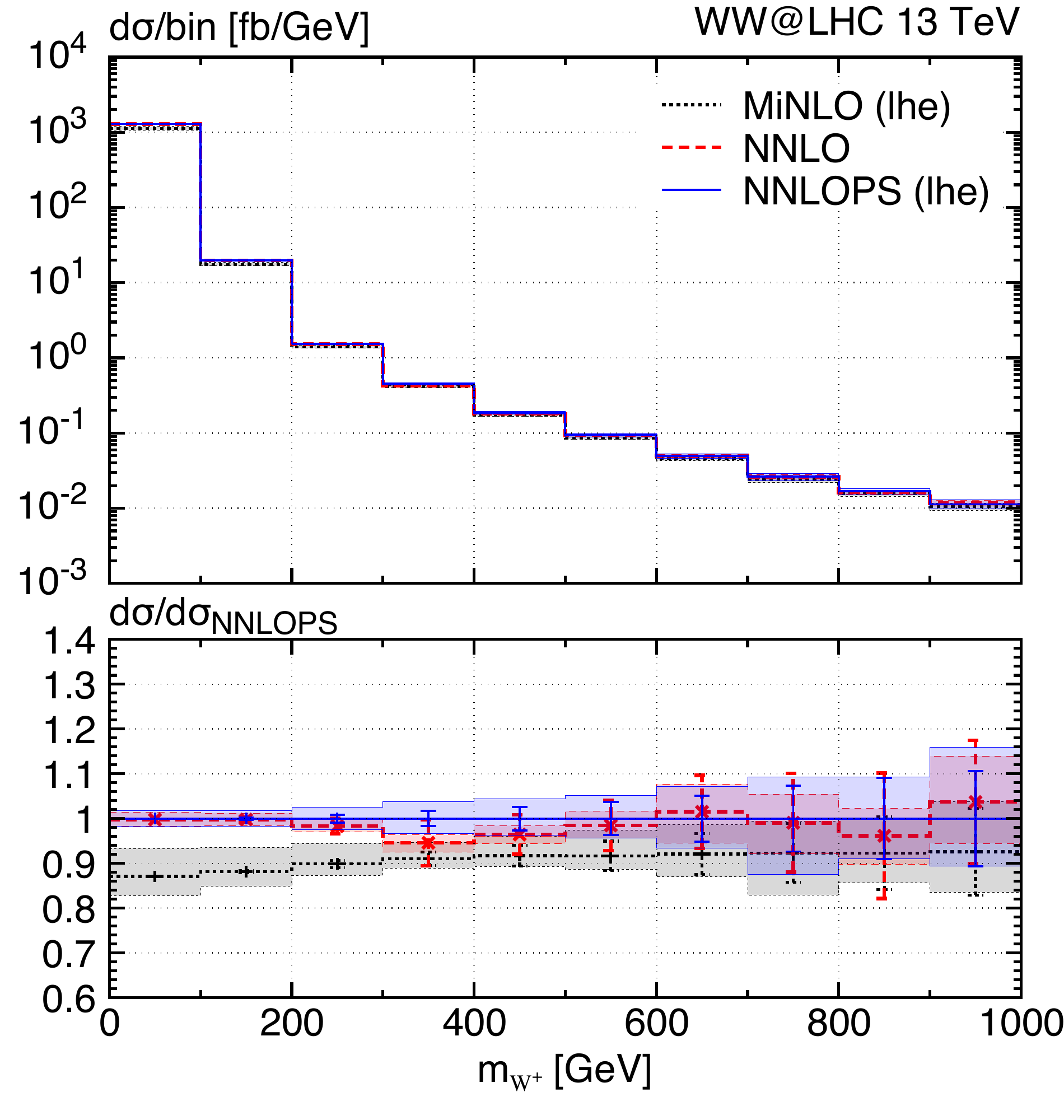}
  \caption{Transverse momenta of the $W^-$ boson (left panel) and
    invariant mass of the leptonic pair coming from the $W^+$ boson
    (right panel), at the $\sqrt{S}=13$ TeV LHC. The red curve is the
    NNLO prediction (central value with uncertainty), the blue and
    black ones are the NNLOPS and \MiNLO{} results, respectively (at
    LHE level, \emph{i.e.} before parton showering). Figures taken from
    ref.~\cite{Re:2018vac}.}
  \label{fig:validation}
\end{figure}
In Figure~\ref{fig:validation} we show two validation plots for
$\ptwm$ and $\mwp$, where we compare the NNLOPS (and {\tt WWJ-MiNLO'})
results (before showering) against the NNLO computation, without any
fiducial cut. We find excellent agreement between NNLOPS and NNLO,
both for an observable used to perform the reweighting ($\ptwm$, left
plot), as well as for $\mwp$, whose dependence was neglected in the
computation of ${\cal W}(\PSb)$: indeed the right plot shows that we
exactly reproduce the NNLO result for $\mwp$ in the peak region, but
also extremely well also quite far in the tail of the
distribution. Several other similar results, supporting the validity
of the approximations we made, can be found in ref.~\cite{Re:2018vac}.

\begin{figure}
  \includegraphics[width=0.49\textwidth]{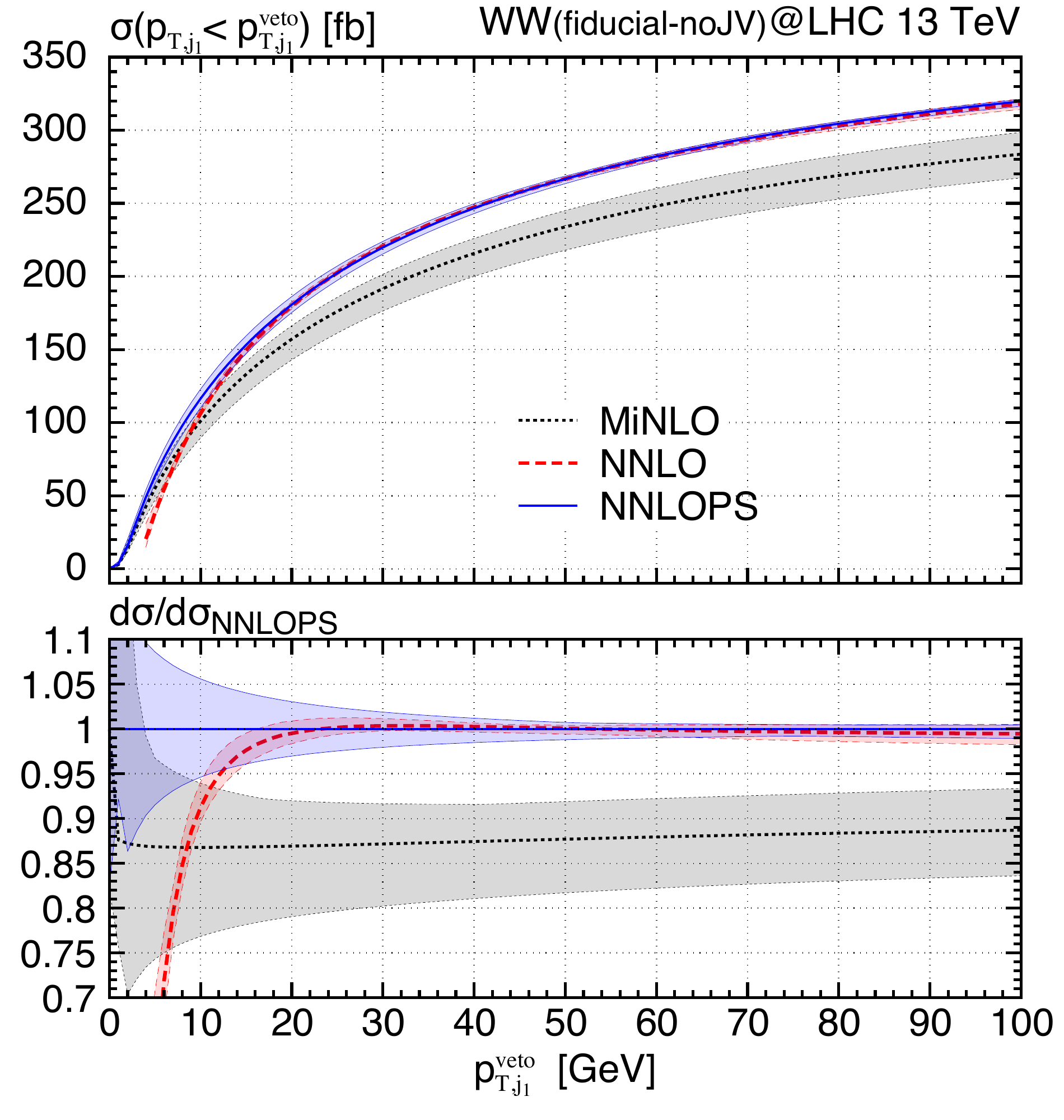}~
  \includegraphics[width=0.49\textwidth]{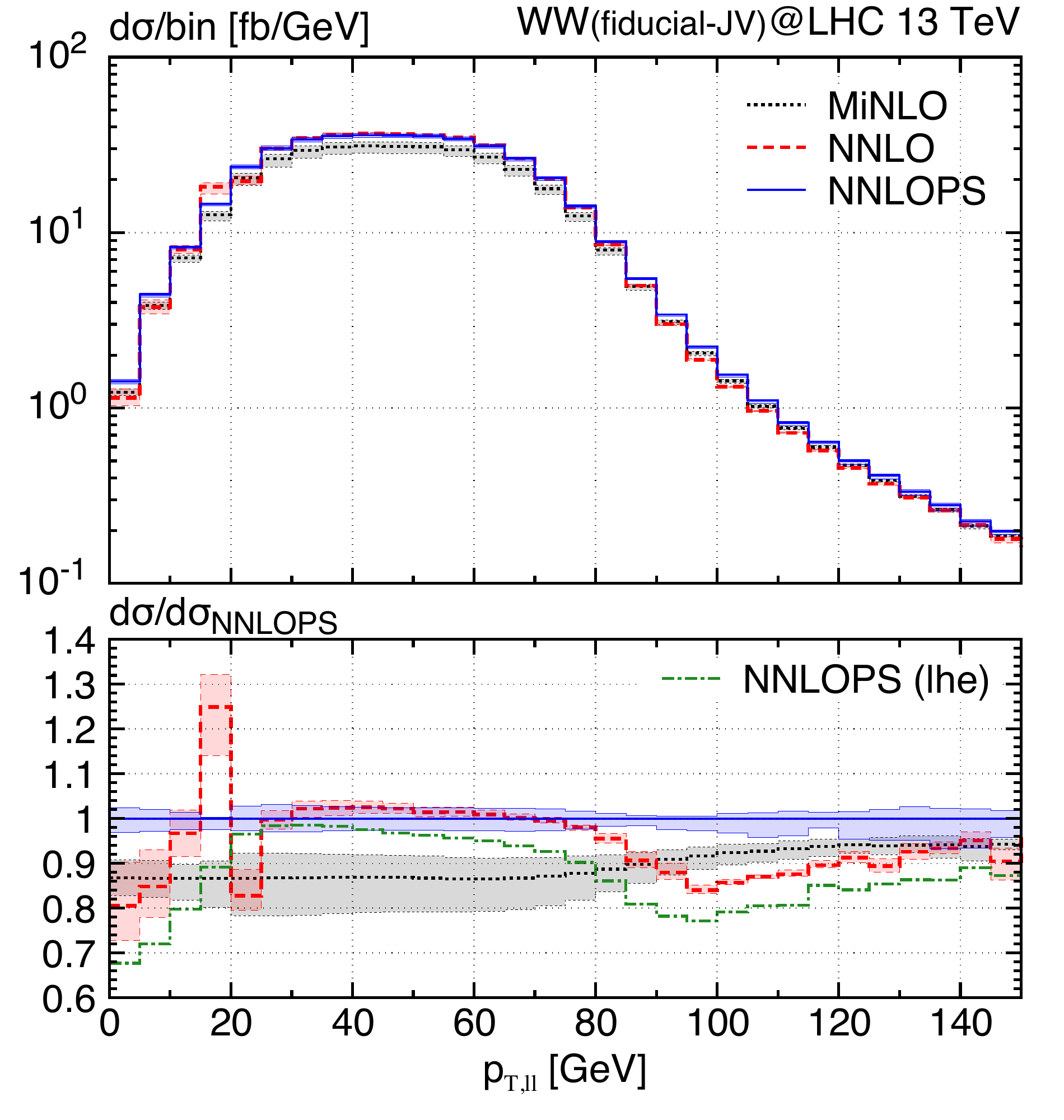}
  \caption{Jet-vetoed cross section (left panel) and transverse
    momentum of the dilepton system (right panel), at the
    $\sqrt{S}=13$ TeV LHC, with fiducial cuts. The red curve is the
    NNLO prediction, the blue and black ones are the NNLOPS and
    \MiNLO{} results after showering, the green one is the NNLOPS
    result before parton showering. Figures taken from
    ref.~\cite{Re:2018vac}.}
  \label{fig:pheno}
\end{figure}
Figure~\ref{fig:pheno} shows instead results where the {\tt
  WWJ-MiNLO'} and NNLOPS events have been showered. Here fiducial cuts
are also applied.\footnote{We refer to ref.~\cite{Re:2018vac} for the
  details.}  In the left panel one can observe the jet-vetoed cross
section, defined as $\sigma(\ptjetone < \ptjetveto) =
\int_0^{\ptjetveto} \dd\ptjetone\,{\dd\sigma}/{\dd\ptjetone}$. The
plot shows that the NNLO provides a good description of the jet veto
down to $\sim 20$\, GeV, supporting the evidence already found in
earlier studies. As expected, for lower values of the jet veto the
NNLO result becomes unphysical, whereas the NNLOPS one remains
well-behaved, due to the all-order resummation of logarithms of the
type $\log(M_{WW}/\ptjetveto)$.  Finally, the right panel of
Fig.~\ref{fig:pheno} displays $\ptll$, the transverse momentum of the
dilepton system. At $\ptll\sim 20$\, GeV the NNLO curve develops a
perturbative instability, due to the presence of a ``Sudakov
shoulder''~\cite{Catani:1997xc} caused by the fiducial cut $\ptmiss{}
> 20$\, GeV. This instability is cured in the NNLOPS (and {\tt
  WWJ-MiNLO'}) results. Furthermore, around $\ptll{}=100$\,GeV a dip
appears in the ratio to the showered NNLOPS prediction: this is due to
the fact that recoiling effects due to the parton shower can cause a
migration of events from one bin to another, and, in some cases, this
can also affect leptonic observables.

The computation presented above is now publicly available within the
\POWHEGBOX{} framework,\footnote{http://powhegbox.mib.infn.it/} and can be used to simulate $W$-boson pair
production fully-exclusively at NNLOPS accuracy. It will be
interesting to further improve this generator by including the effect
of $gg$-induced contributions (for which an NLO+PS study, for the $ZZ$
case, was performed in ref.~\cite{Alioli:2016xab}), as well as to
reach NNLOPS accuracy without the need of an explicit reweighting.

\end{document}